\documentclass[11pt, letterpaper]{article}
\usepackage[margin=1in]{geometry}
\usepackage[utf8]{inputenc} 
\usepackage{siunitx}

\usepackage[sort,compress]{natbib}
\setcitestyle{numbers}

\usepackage[toc,page]{appendix}

\usepackage[format=plain,labelfont={bf},textfont=it]{caption}
\usepackage{graphicx}
\usepackage{float}

\usepackage[table,xcdraw]{xcolor}
\usepackage{comment}

\usepackage{totcount}
\newtotcounter{citnum} 
\def\oldbibitem{} \let\oldbibitem=\bibitem
\def\bibitem{\stepcounter{citnum}\oldbibitem}

\title{Reevaluating the Role of Race and Ethnicity in Diabetes Screening}

\author{
  Madison Coots \thanks{Correspondence may be sent to Madison Coots at \texttt{mcoots@g.harvard.edu}.} \\
  Harvard University
  \and
  Soroush Saghafian\\
  Harvard University
  \and
  David Kent\\
  Tufts University
  \and
  Sharad Goel\\
  Harvard University
}

\date{}

\begin{document}
\maketitle

\thispagestyle{empty}
\setlength{\fboxsep}{1em}

\noindent\fbox{\begin{minipage}{.95\textwidth}
\parbox{38em}{
\textbf{\large Key Messages}
\begin{itemize}
    \item There is active debate over whether to consider patient race and ethnicity when estimating disease risk. By accounting for race and ethnicity, it is possible to improve the accuracy of risk predictions, but there is concern that their use may encourage a racialized view of medicine.
    \item In diabetes risk models, despite substantial gains in statistical accuracy from using race and ethnicity, the gains in clinical utility are surprisingly modest.
    \item These modest clinical gains stem from two empirical patterns: first, the vast majority of individuals receive the same screening recommendation regardless of whether race or ethnicity are included in risk models; and second, for those who do receive different screening recommendations, the difference in utility between screening and not screening is relatively small.
    \item Our results are based on broad statistical principles, and so are likely to generalize to many other risk-based clinical decisions.
\end{itemize}}
\end{minipage}}

\clearpage
\section*{Introduction}
\label{intro}

Approximately 1 in 10 Americans suffers from Type 2 diabetes. Diabetes can lead to several serious health problems, such as heart disease, kidney disease, and vision loss. But, if detected early, patients can receive pharmacotherapy and make corrective changes to their diet and lifestyle to better manage their health. Although every individual could be regularly screened for diabetes, screening comes with monetary and non-monetary costs (e.g., taking time off from work may result in lost income). Consequently, the medical community recommends that only those with at least a moderate risk of developing diabetes undergo screening. Results by Aggarwal et al. suggest that individuals typically benefit from screening if their risk of diabetes is above 1.5\%~\citep{agarwal2022diabetes}. To follow this guidance, statistical algorithms can be used to predict the diabetes risk for each individual, recommending screening for those with predicted risk above the threshold.

These risk predictions are often produced using variables such as age and body mass index (BMI). There is debate over whether to additionally base risk predictions on an individual’s race and ethnicity to account for observed disparities in diabetes incidence rates across  demographic subgroups in the United States.~\citep{agarwal2022diabetes} Past work has argued that including race and ethnicity substantially improves the accuracy of clinical prediction models, and that their omission could exacerbate disparities in health outcomes~\citep{manski2022patient, manski2022using}. However, there is persistent concern over the use of race and ethnicity in estimating disease risk---for diabetes and beyond---and leading journals have published articles critical of “race-aware” predictions~\citep{vyas2020hidden, cerdena2020race, basu2023use}. Including race and ethnicity as inputs to predictive models may, for instance, inadvertently reinforce pernicious attitudes of biological determinism or lead to greater stigmatization of individuals who are already marginalized. In part for these reasons, several hospitals have recently moved away from reporting race-adjusted glomerular filtration rate estimates, instead reporting a race-unaware value, both to avoid race-based predictions and to mitigate concerns that a race-aware model may deprioritize Black patients for kidney transplantation ~\citep{powe2020black, diao2021clinical, eneanya2019reconsidering}.

We offer a new perspective on the statistical and clinical utility of race and ethnicity in diabetes risk estimation. We begin by showing that racial and ethnic minorities---particularly Asian Americans---are substantially more likely to develop diabetes than White Americans with comparable age and BMI, in line with past analyses. Accordingly, considering race and ethnicity can substantially improve diabetes risk predictions for both White and non-White individuals. However, by adopting a utility-based decision framework, we then show that these greatly improved predictions do not result in the commensurately large clinical benefits many might expect. This finding stems from two patterns in the data: first, while the more accurate race-aware model changes predictions for all patients, decisions (in terms of screening recommendations) change for only a small fraction of patients; second, among those who do receive different screening recommendations, the net value of screening is relatively small since their risk of diabetes is close to the decision threshold.

Our analysis is based on publicly available data from the National Health and Nutrition Examination Survey (NHANES)~\citep{nhanes}. NHANES combines interview responses with laboratory data to provide insight into the health and nutritional status of adults and children in the U.S. The survey is conducted every two years by the National Center for Health Statistics and is frequently used by researchers to assess the prevalence of major diseases and their risk factors across the U.S. population. We use the four NHANES cycles from 2011-2018. Following Aggarwal et al., we restricted our sample to approximately 18,000 patients who were not pregnant, were 18--70 years old, and had a BMI between 18.5 and 50.0 \si[per-mode=symbol]{\kilogram\per\meter^2}.

\section*{Results}
\begin{figure}[t!]
	\centering
	\includegraphics[width=.9\columnwidth]{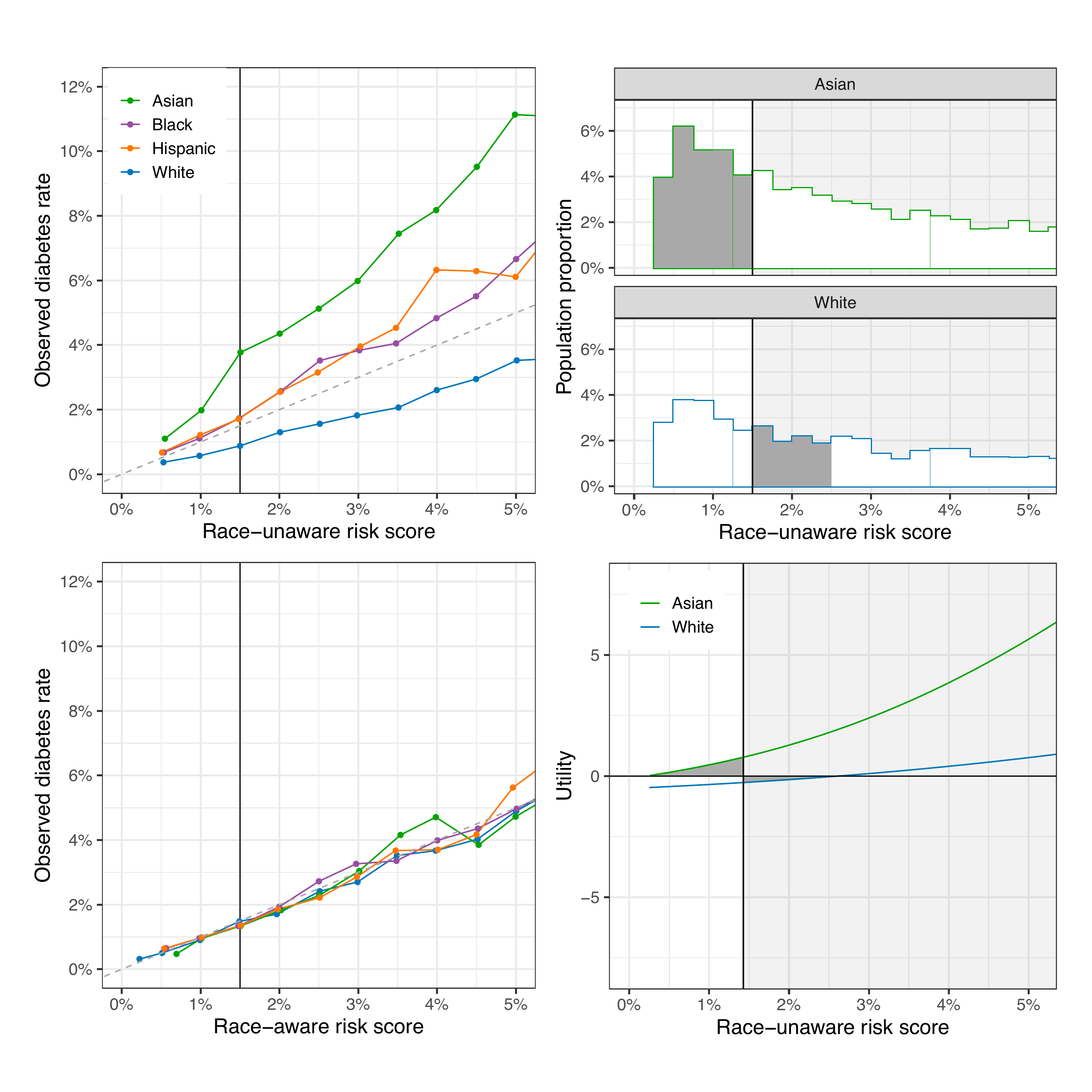}
	\caption{Assessing the statistical and clinical utility of race and ethnicity in diabetes risk estimation. Upper left: Calibration plot for a race-unaware risk model, showing that racial minorities have significantly higher empirical risk of diabetes compared to their race-unaware risk prediction; Lower left: Calibration plot for a race-aware risk model, showing that predicted and empirical risk are well-aligned across groups; Upper right: Distribution of race-unaware predicted risk, highlighting individuals for whom the race-unaware screening decision is at odds with their empirical utility for being screened. Lower right: Plot showing utility as a function of race-unaware risk predictions, indicating that individuals who receive incorrect screening recommendations incur relatively small utility loss (highlighted in grey). In all four panels, the horizontal axis excludes those with higher than 5\% predicted risk to highlight the region near the decision threshold.
}
	\label{fig:main}
\end{figure}

\paragraph{Race-unaware predictions of diabetes risk are substantially miscalibrated by race and ethnicity.}
Among the subset of individuals with a “race-unaware” estimated diabetes risk of 1\%---based on age and BMI---about 1\% were in reality found to have diabetes. But when we disaggregate this group of individuals by race and ethnicity, we see that the model systematically underestimates diabetes risk for Asian, Black, and Hispanic patients, while overestimating diabetes risk for White patients. In particular, Asian patients with an estimated risk of 1\% under the race-unaware risk prediction model in reality have diabetes at a rate of approximately 2\%, double the nominal estimate. That is, of the approximately one million Asian American adults in the U.S. with a race-unaware predicted diabetes risk of 1\%, about 2\% of them in reality would be found to have diabetes. Similarly, White patients with a predicted risk of 2\% under the race-unaware model have diabetes at a rate just under 1.5\%. These miscalibrated predictions lead to misclassifications: with a screening threshold of 1.5\%---in line with current guidelines---the race-unaware model would incorrectly fail to recommend screening for some Asian American patients who have relatively high risk of diabetes and would incorrectly recommend screening for some White patients who have relatively low risk of diabetes. The upper-left plot of Figure~\ref{fig:main} compares the empirical rate of diabetes to the race-unaware risk predictions across the risk spectrum, illustrating the miscalibration of predicted risks by race and ethnicity.

\paragraph{Race-aware models improve risk predictions.}
One way to correct the observed miscalibration is to incorporate race and ethnicity into risk predictions. The lower-left plot of Figure~\ref{fig:main} compares the empirical rate of diabetes to race-aware predicted risks, based on age, BMI, and race and ethnicity. The predicted race-aware risks are largely in line with empirical diabetes rates across groups.
Further, it seems hard to rectify the miscalibration of race-unaware models without explicitly considering race and ethnicity, as the miscalibration persists for race-unaware models that consider a variety of additional factors beyond age and BMI, including family history (see Figure~\ref{fig:extended_calibration} in the Appendix for further details).

\paragraph{Despite improved accuracy, race-aware predictions yield limited net benefits.}
Race-aware predicted risk scores greatly improve the calibration of estimates, improving aggregate outcomes for all race and ethnicity groups. However, perhaps surprisingly, the overall clinical benefits are not as large as one might expect. To quantify the added value of race-aware risk scores, we adopt a utility-based framework where both costs and gains in health are expressed in dollars by specifying an exchange rate. We begin by assuming a constant cost for screening (which we normalize to equal 1) and a constant benefit $r$ for detecting diabetes, which captures the net benefit of early detection and treatment of the disease. The screening cost encapsulates a wide range of monetary and non-monetary considerations, such as the cost to perform the test and the cost of taking time away from work to be screened. Based on this framework, the optimal policy is to screen patients if and only if their predicted risk of diabetes exceeds $t=1/r$, which represents the ``point of indifference.'' In other words, the expected benefits of screening exceed the costs if, and only if, one’s risk of diabetes is greater than $1/r$. For the purposes of our analysis, we set the benefit $r$ for detecting diabetes to 70, consistent with guidelines that suggest a screening threshold for diabetes of approximately 1.5\%. (In Figure~\ref{fig:sensitivity_analysis} in the Appendix, we show that our results are qualitatively similar across a wide range of choices for $r$.) 

Across the entire patient population, we estimate the per capita utility gain from the race-aware model is 0.015---where we recall that the cost of screening is normalized to equal 1. If we assume the cost of screening equals \$100, then the race-aware diabetes risk model yields an average per capita gain of approximately \$1.50 over a race-unaware model. 

\begin{table}[t!]
\centering
\begin{tabular}{cc}
\textbf{Race and ethnicity} & \textbf{Per capita utility gain} \\ 
\hline
Asian              & 0.077                   \\
White              & 0.016                   \\
Hispanic           & 0.0005                  \\
Black              & 0.0004                 
\end{tabular}
\caption{The per capita utility gain of a race-aware risk model over a race-unaware model, disaggregated by race and ethnicity.}
\label{tab:utility}
\end{table}

In Table~\ref{tab:utility}, we show the results of this utility analysis by race and ethnicity. Asian individuals experience the greatest utility gains from a race-aware model, with a gain of about \$7.77 per person on average. White individuals follow with a utility gain of approximately \$1.60 per person (from the avoidance of some unnecessary screening), and Black and Hispanic individuals experience near-zero utility gains from the use of a race-aware model over a race-unaware model. The largest gains accrue to Asian Americans in part because, as discussed above, the race-unaware model most significantly underestimates diabetes risk for this group.

Given that the race-unaware risk model is starkly miscalibrated, it is perhaps surprising that the race-aware model does not yield larger utility gains. Two factors help explain this phenomenon. First, most individuals receive the same screening recommendation under a race-aware model as under a race-unaware model, as recommendations only change for the relatively few individuals close to the decision threshold. Specifically, 84\% of Asian individuals, 93\% of White individuals, 96\% of Hispanic individuals, and 97\% of Black individuals receive the same screening decision under both models. The vast majority of  individuals thus accrue no gains from the race-aware model. Second, the small number of patients near the threshold who do receive different screening recommendations under the two risk models are relatively indifferent between being screened and not being screened---precisely because they are close to the screening threshold---and so accrue small utility gains.

The upper-right and lower-right plots in Figure~\ref{fig:main} illustrate these patterns. The upper-right plot shows the distribution of race-unaware risk scores for Asian and White individuals, highlighting the relatively small number of patients for whom the screening recommendation is at odds with their empirical utility for being screened. The lower-right plot shows the empirical utility of screening as a function of race-unaware risk score, for both Asian and White individuals. Asian individuals with nominal race-unaware diabetes risk lower than the 1.5\% screening threshold would be counselled against screening even though they have positive utility for screening. Analogously, some White individuals with race-unaware risk estimates above the screening threshold would be counselled in favor of screening even though they have negative utility for screening. But in both cases, the loss of utility for these errors is relatively small. 

\paragraph{Generalizing these results to other decisions.}
Because these results are based on broad principles, we believe they are likely to generalize to many other settings in which race-aware models might improve predictions. The key insight here is that when the decision threshold is determined by the “point of indifference” (i.e., where the costs and benefits of an intervention are exactly balanced), decisions are likely to change only for patients who are close to this “point of indifference”, thus limiting the value of improved classification. Where risk prediction is used in a shared decision-making context, the decision threshold is generally determined by that point of indifference. 

\section*{Discussion}
The principles illustrated here might not apply under conditions of scarcity, where rationing of resources may yield thresholds far from the point of indifference (e.g., organ transplantation). Consider an example where severe budgetary constraints mean that only the riskiest 50\% of Americans could be screened for diabetes, which corresponds approximately to a 6\% screening threshold. If the “point of indifference” (from the patient perspective) remained at 1.5\%, the marginal benefit of using a race-aware over a race-unaware approach for targeting would improve utility for racial minorities by an order of magnitude.

Further, our analysis is contingent on the specific utility function that we use to evaluate screening decisions. For example, we implicitly assume that the value of detecting diabetes is comparable across race and ethnicity groups. In settings where that is not the case, the benefits of a race-aware approach are likely larger than what we find here~\citep{thomas2021race}. Second, we assume that the benefit of detecting diabetes is independent of an individual’s age. A more detailed analysis might explicitly consider quality-adjusted life years. Third, to aid interpretation, we estimate that the monetary and non-monetary costs of screening are approximately \$100 (i.e., that one util equals \$100). Given the subjective nature of some of the screening costs (e.g., an individual’s value for time), accurately estimating this quantity is challenging. Finally, we have primarily considered per capita utility gains, but one could alternatively consider aggregate population-level utility benefits, which are considerably larger. Nonetheless, we believe our simplified analysis helps illustrate that substantial gains in statistical accuracy may still result in modest gains in clinical utility.

Race-aware risk models can improve estimates of diabetes risk, but the clinical value of considering race and ethnicity is smaller than the improvement in predictions might suggest. Thus, if there are important reasons to avoid consideration of race and ethnicity in risk prediction---for example, from decreased trust in the medical system---the clinical benefits may not justify their use. Our results likely extend to a variety of contexts in medicine and beyond where the explicit use of race and ethnicity in predictive models is contested, although the specifics in each case need to be considered. We hope that our analytic approach helps researchers and policymakers better understand and balance these underlying trade-offs.

\section*{Contributors and sources}

Madison Coots is a Ph.D. student in public policy. Dr. Saghafian is an associate professor of public policy with expertise in the development and application of methods in operations research and management science to examine societal problems in healthcare. Dr. Kent is a general internist and researcher with two decades of experience in clinical prediction modelling. Dr. Goel is a professor of public policy with expertise in algorithmic fairness and computational approaches to public policy. Coots wrote the initial draft of the manuscript and is the guarantor. All authors reviewed and revised the manuscript.

\section*{Data and code availability}
Data and code to reproduce our analysis are available at: https://github.com/madisoncoots/race-in-diabetes-screening.

\section*{Acknowledgements}
Dr. Kent is funded through a ``Making a Difference'' and Presidential Awards from The Greenwall Foundation.

\clearpage
\bibliography{refs}

\clearpage
\appendix
\renewcommand\thefigure{\thesection\arabic{figure}}    
\section{Appendix}
\setcounter{figure}{0}    


\begin{figure}[H]
\centering
\textbf{Calibration plot for an extended race-unaware diabetes risk model}\par\medskip
	\includegraphics[width=0.45\columnwidth]{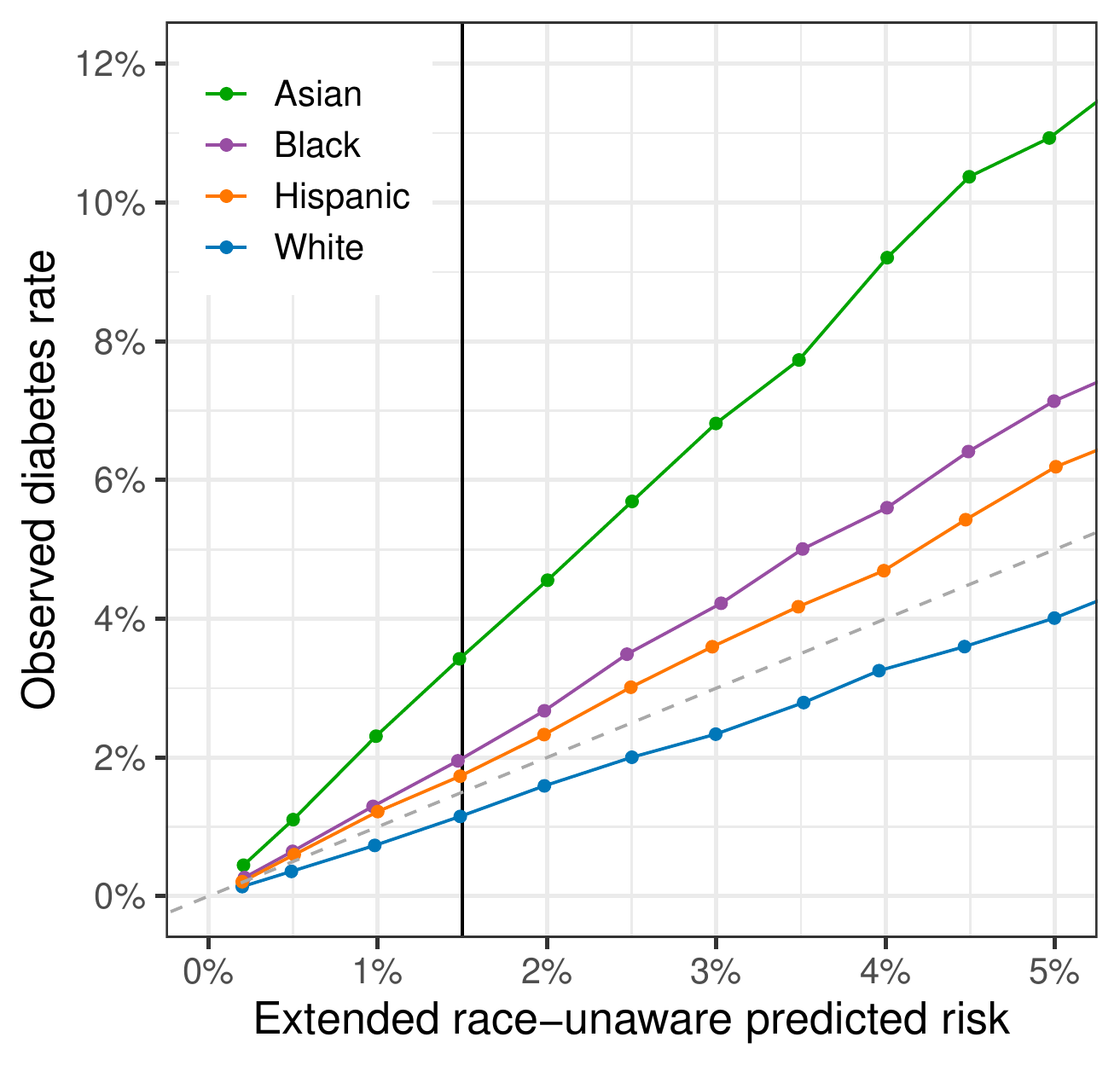}
	\vskip 0.1in
	\caption{Calibration plot for an extended race-unaware risk model that includes the following covariates in addition to age and BMI: gender, weight, height, waist circumference, self-reported greatest weight, whether a patient’s close family members have diabetes, whether the patient is depressed, income, health insurance status, and whether the patient feels food secure. 
}
	\label{fig:extended_calibration}
\end{figure}


\begin{figure}[H]
\centering
\textbf{Sensitivity analysis over the reward for detecting diabetes}\par\medskip
	\includegraphics[width=0.45\columnwidth]{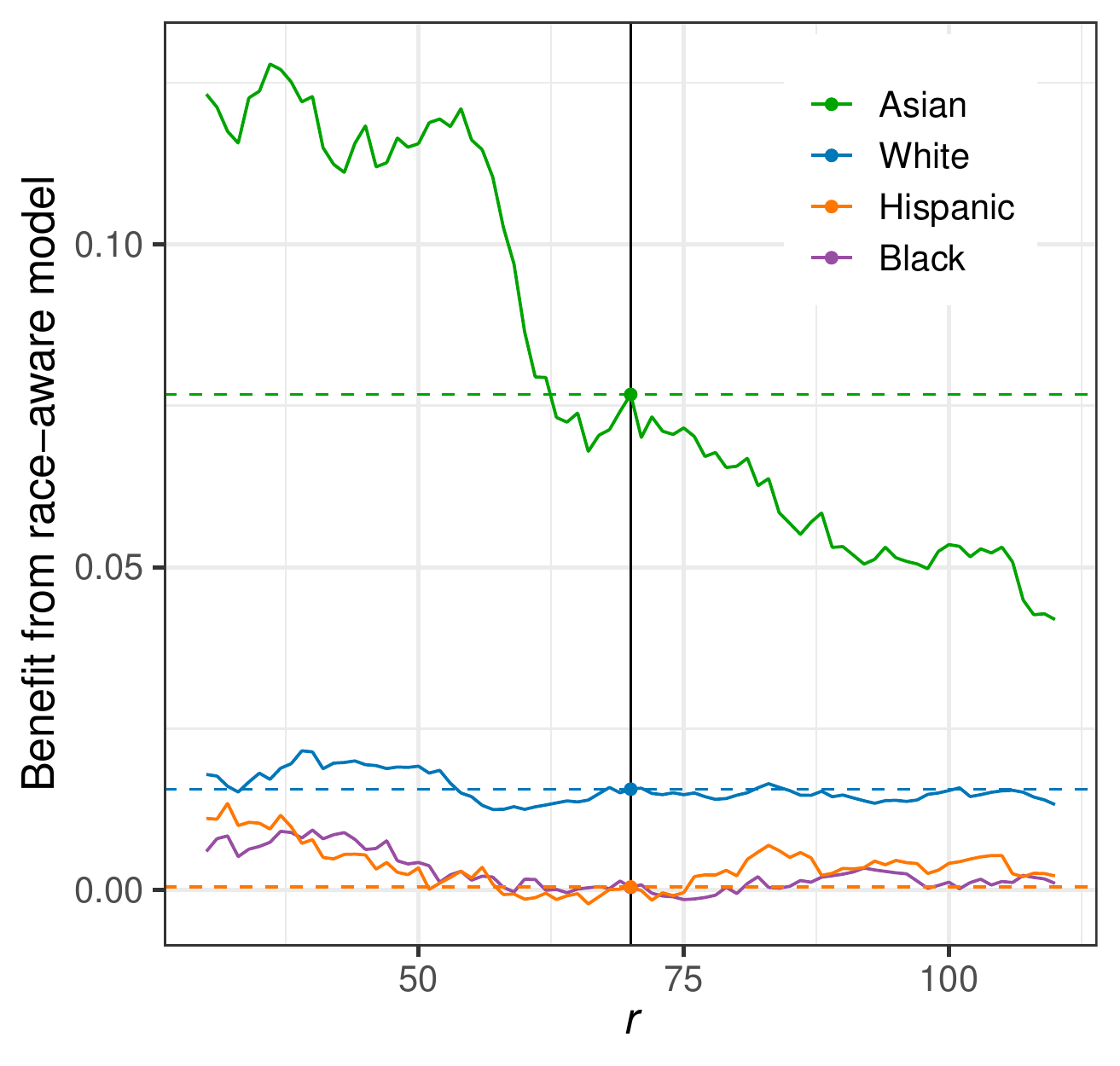}
	\vskip 0.1in
	\caption{When varying the reward for diabetes detection r, we observe relatively small changes in the benefit from the use of a race-aware risk model across groups.  
}
	\label{fig:sensitivity_analysis}
\end{figure}


\begin{figure}[H]
\centering
\textbf{Subgroup utility plots}\par\medskip
	\includegraphics[width=0.45\columnwidth]{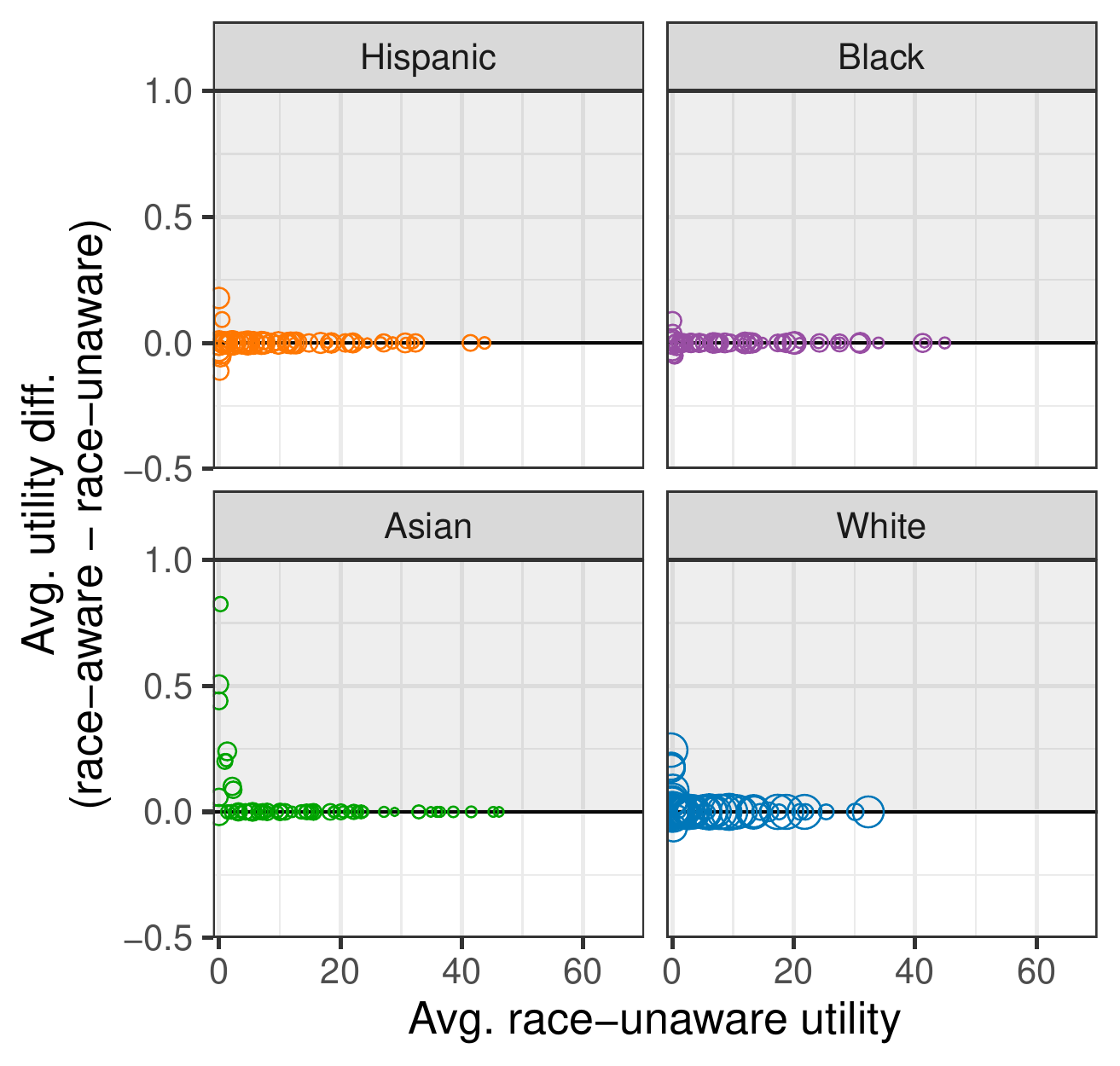}
	\vskip 0.1in
	\caption{Each circle represents a different patient subgroup by age, BMI, and race, and the circles are sized according to the number of patients in that subgroup. }

	\label{fig:subgroup}
\end{figure}



\end{document}